# On the Transition of a Non-Equilibrium System to an Equilibrium System

E. G. D. Cohen

*Rockefeller University, 1230 York Avenue, New York, NY 10065 USA,*

*Department of Physics and Astronomy, The University of Iowa, Iowa City, IA, 52242*

March 26, 2015

Abstract

It is shown that the most important feature of Non-Equilibrium Thermodynamics is not the entropy production, but the organization of the currents in order to flow. This is also needed to obtain the maximum entropy in the equilibrium state, as is required by Equilibrium Thermodynamics.

It is a real privilege for me to contribute to this Festschrift in honor of Jacques Yvon. In 1969, more than 45 years ago, he wrote a book about entropy: "Correlations and Entropy in Statistical Mechanics" [1]. He was clearly an original, who thought for himself. In his book he uses a cluster-like expansion of the entropy, in order to obtain in a systematic way its space correlations.

In this paper I also want to write about entropy, as Yvon did, but I will discuss a different aspect of the equilibrium entropy, which leads to a new insight into this quantity.

---

[1] In this book also the so-called BBGKY (Bogolubov-Born-Green-Kirkwood-Yvon) hierarchy – a name coined by G. E. Uhlenbeck – can be found.

The entropy was introduced into physics by Rudolf Clausius. It took him from 1854-1865 to straighten out this new concept to his satisfaction [1].

I want to discuss here the following question.

In a non-equilibrium system, according to Non-Equilibrium Thermodynamics, there are flowing currents, which must have dissipation or entropy production, according the Second Law of Thermodynamics' impossibility of a *perpetuum mobile* of the second kind. Otherwise one would have a machine, which would function eternally by absorbing heat from an (infinite) heat reservoir and using this heat completely to run this machine.

In order to create a non-equilibrium system from an equilibrium system, appropriate external forces have to be applied to the equilibrium system. Thus, e.g. to obtain from a system A in thermal equilibrium at a temperature $T_A$, a non-equilibrium system with a heat flux, one attaches to it two heat reservoirs of different temperatures $T_B > T_C$ (cf figure 1). This temperature difference will create a temperature gradient in the original system A, leading to a heat current $\mathbf{J}_A$, from the high temperature $T_B$ to the low temperature $T_C$ heat reservoir. It will be assumed that a laminar flow takes place [2].

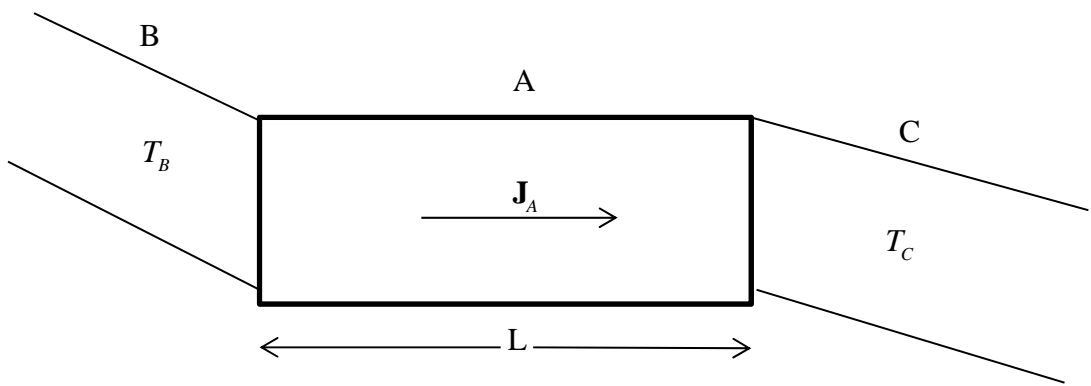

**Figure 1.** A heat current $\mathbf{J}_A$ in a fluid A flows from a heat reservoir B – at a higher temperature $T_B$ – via the fluid A of length $L$, to a heat reservoir C – at a lower temperature $T_C$ – due to a temperature gradient $(T_B - T_C)/L$.

In order for the system A to return to its original (isolated) equilibrium state, one has to remove the two heat reservoirs, so that the temperature gradient will gradually disappear and the heat current with it.

In books on Non-equilibrium Thermodynamics, e.g. the standard classical book "Non-Equilibrium Thermodynamics" by S. R. De Groot and P. Mazur [3], the main emphasis is that the current must have dissipation, or entropy production, as mentioned above.

A problem then arises in that there seems to be no source to provide the entropy of the equilibrium system, when a non-equilibrium system goes to an equilibrium state. This, since the only source for the equilibrium entropy would be the non-equilibrium entropy production, which disappears – together with the heat current – in the equilibrium state.

The answer to this difficulty is the following: in order to flow in a certain direction, the current $\mathbf{J}_A$ needs an ordering or organization of the fluid particles.

In Non-equilibrium Thermodynamics books, in particular in the book of S. R. De Groot and P. Mazur [3], only the disappearance of the entropy production is mentioned, when the system reaches the equilibrium state, but there is then *no* mechanism is given to provide for the equilibrium entropy.

Therefore, the question arises: what *is* the source of the equilibrium entropy, when the system has reached the equilibrium state?

The answer is that it has been overlooked, that the heat current $\mathbf{J}_A$ must necessarily have an (inner) ordering or *organization*[2], since the particles have to flow in a certain direction. This is a direct *physical* consequence of a mathematical representation as a vector, with a magnitude, as well as a direction. Then, when the heat baths are removed and the gradient gradually disappears, the heat current will more and more disintegrate and randomize, losing its organization. When the equilibrium state is reached, a maximum randomization or entropy will have been obtained, as required by Equilibrium Thermodynamics.

Therefore, it is the current's organization, which disappears when the system reaches equilibrium and then provides the ensuing randomization of the fluid particles to produce the equilibrium entropy. This implies that the Non-Equilibrium Thermodynamics [3] textbooks, where the current is merely represented *mathematically* by a symbol $\mathbf{J}$, has in addition a *physical* nature, in that it is an organization of the fluid particles, above and beyond just a mathematical symbol.

When the external forces, which created from an equilibrium system a non-equilibrium system are removed, the temperature gradient as well as the heat current will gradually disintegrate and disappear, when an equilibrium state will be reached. Therefore, it is the loss of the current's organization, which leads to an increasing randomization of the system, and creates the equilibrium system's maximum entropy, in agreement with Equilibrium Thermodynamics.

---

[2] The organization of the current can be observed a.o. by inserting a dye into the fluid, which will show the streamlines of the organized currents, as shown in [4].

Thus, one could call the current's organization "antropy", from anti-entropy, in view of the transmutation of the non-equilibrium current organization into the equilibrium entropy.

Alternatively, when a temperature gradient is introduced into a system in equilibrium, a non-equilibrium heat flow will appear, which leads to an organized heat current. The above suggests that it is important to consider, together with the non-equilibrium state, *also* the equilibrium state and not only consider the disappearance of the entropy production of the non-equilibrium state as in [3].

In addition, in Non-Equilibrium Thermodynamics textbooks as in the standard classical book [3], where the current is represented mathematically by a vector, $\mathbf{J}$, which has an additional important physical aspect, as being *organized*, in order for the fluid particles to flow in a certain direction.

The reason that the current's organization is not mentioned in [3] is that in non-equilibrium thermodynamics textbooks, only the non-equilibrium state itself is considered and the equilibrium state is only mentioned in connection with the disappearance of the entropy production.[3]

Finally, the difference between mathematics and physics has, in my opinion, nowhere been better characterized by J.W. Gibbs in a footnote on page 35 of his classical book 1902:

---

[3] This reminds me of my postdoctoral stay with Professor G. E. Uhlenbeck, who occasionally mentioned derogatorily that someone had become an "expert" working in one field only. Although I then aspired just that, later I realized the wisdom of his remark.

"Elementary Principles in Statistical Mechanics" Yale University Press, (1902), Dover Publications, New York (1960). There he says in a footnote: "…. Again two attracting particles should be able to do an infinite amount of work in passing from one configuration (which is regarded as possible) to another, is a notion which, although perfectly intelligible in a mathematical formula, is quite foreign to our ordinary conceptions of [ponderable] matter [i.e. to physics]."